\documentclass[10pt,journal,compsoc]{IEEEtran}
\ifCLASSOPTIONcompsoc
  \usepackage[nocompress]{cite}
\else
  \usepackage{cite}
\fi

\usepackage{algorithm}
\usepackage{algorithmic}
\usepackage{caption}
\usepackage{subcaption}
\usepackage{authblk}
\usepackage{comment}
\usepackage{setspace}
\usepackage{amsmath,amssymb,graphicx,epstopdf,textpos,rotating}
\usepackage{amsmath,amssymb}
\usepackage{color}
\usepackage{caption}
\usepackage{todonotes}
\usepackage{subcaption}
\makeatletter
\def\blfootnote{\xdef\@thefnmark{}\@footnotetext}
\makeatother
\usepackage{hyperref}

\usepackage{color}

\newcommand{\bp}{{\bf p}}

\newcommand{\N}{\mathbb{N}}
\newcommand{\R}{\mathbb{R}}
\newcommand{\bG}{{\mathbb{G}}}
\newcommand{\bEg}{{\mathbb{E}}}

\newcommand{\bP}[1]{{{\bf P}}\left[{#1}\right]}

\newcommand{\bE}[1]{{\mathbb{E}}\left[{#1}\right]}
\newcommand{\1}[1]{{\bf 1}\left[#1\right]}

\newtheorem{theorem}{Theorem}[section]

\newtheorem{lemma}[theorem]{Lemma}

\newcommand{\fsquare}{\vrule height6pt width7pt depth1pt}   
\newcommand{\myproof}{{\hfill \\ \bf Proof. \ }}           
\newcommand{\myendpf}{\hfill\fsquare \\[0.1in]}

\allowdisplaybreaks

\title{Dynamics of node influence in \\network growth models\thanks{This document does not contain technology or technical data controlled under either the U.S. International Traffic in Arms Regulations or the U.S. Export
Administration Regulations.}}

\author{\IEEEauthorblockN{Shravika Mittal\textsuperscript{$\star$},
        Tanmoy Chakraborty\textsuperscript{$\star$}}, Siddharth Pal\textsuperscript{$\dagger$}\\
\IEEEauthorblockA{\textit{\textsuperscript{$\star$}Dept. of CSE, IIIT-Delhi, India \textsuperscript{$\dagger$}Raytheon BBN Technologies, USA}\\
\{shravika16093, tanmoy\}@iiitd.ac.in, siddharth.pal@raytheon.com}
        }

\date{\vspace{-5ex}}   

\begin{document}

\IEEEtitleabstractindextext{
\begin{abstract}

Many classes of network growth models have been proposed in the literature for capturing real-world complex networks. Existing research primarily focuses on global characteristics of these models, e.g., degree distribution. We aim to shift the focus towards studying the network growth dynamics from the perspective of individual nodes. In this paper, we study how a metric for node influence in network growth models behaves over time as the network evolves. This metric, which we call node visibility, captures the probability of the node to form new connections. First, we conduct an investigation on three popular network growth models -- preferential attachment, additive, and multiplicative fitness models; and primarily look into the ``influential nodes'' or ``leaders” to understand how their visibility evolves over time. Subsequently, we consider a generic fitness model and observe that the multiplicative model strikes a balance between allowing influential nodes to maintain their visibility, while at the same time making it possible for new nodes to gain visibility in the network. Finally, we observe that a spatial growth model with multiplicative fitness can curtail the global reach of influential nodes, thereby allowing the emergence of a multiplicity of ``local leaders" in the network.

\end{abstract}
\begin{IEEEkeywords}
Network growth models, node dynamics, Barabasi-Albert graphs, fitness based models, spatial models.
\end{IEEEkeywords}}

\maketitle

\IEEEdisplaynontitleabstractindextext

\IEEEpeerreviewmaketitle

\section{Introduction}
\label{sec:Introduction}
Over the past two decades, complex networks have been used to model real-world systems across different domains ranging from social, biological, information, and technological systems~\cite{newman2010networks,newman2003SIAM}. 
Investigating the behavior of influential entities or leaders in these real-world networks would help us understand how they are able to gather and maintain prominence over time. 
For instance, influential papers in citation networks continue to acquire new citations every year~\cite{wang2013quantifying,chakraborty2015categorization}. Likewise, celebrities
in online social networks keep increasing their follower count over time. These influential entities act as potential spreaders of information in networks. Therefore, keeping a track of their characteristics in an evolving network could have significance in applications ranging from viral marketing and target advertisement to rumor and epidemic control, and protection from  spam attacks \cite{CHEN20121777}, \cite{Guo_2020}, \cite{10.5555/1619797.1619865}. To understand and model the dynamics of how a leader node maintains its influence over time, one needs to study the temporal behavior of nodes in a network.
In this paper, we introduce a notion, called {\em visibility of a node} which is defined as the {\em probability of the node to form new connections in a growing network}. For instance, in a preferential attachment model~\cite{barabasi_2009,barabasi1999emergence}, the visibility of a node is proportional to its degree, and inversely proportional to the number of edges in the network. An essential aspect of the study is to investigate the {\em visibility profile} of a node which characterizes the temporal evolution of the node's influence as the network grows. We argue that studying the visibility profile of nodes leads to a better understanding of network evolution due to attachment dynamics, which might not be possible to obtain by simply analysing global network properties such as degree distribution or local node-centric properties such as degree, clustering coefficient, etc.
Similar to node persistence over time studied in~\cite{noulas2015topological}, our approach allows to make headway into this understanding by characterizing the {\em visibility behavior of leaders} in the network.  
While the framework is applicable to arbitrary nodes as well, it is more interesting to first understand the leaders' behavior. For example, Chakraborty et al. \cite{chakraborty2015categorization} argued that the growth of the degree  of a node (its visibility) in a citation network follows one of the five patterns -- early rise, late rise, frequent rise, steady rise and steady drop. In a subsequent study \cite{10.1007/s10115-017-1080-y}, they also concluded that  highly-cited papers and authors (leaders) follow steady rising pattern. However, it was not clear whether existing network growth models are able to describe such patterns particularly for leader nodes \cite{mohapatra2020modeling}. This motivates us to study the temporal evolution of the visibility of nodes, in particular leaders in  networks generated by different network growth models.   



We study the visibility of influential nodes\footnote{We use ``leaders'' and ``influential nodes'' interchangeably to denote high-degree nodes that can keep attracting new edges over long periods of time.} in the graphs simulated by the following network growth models. 
Barab{\'a}si-Albert (BA) model~\cite{barabasi1999emergence}, {\em aka} the preferential attachment model, was able to explain power law behavior in real-world networks using the idea of network growth and the ``rich-get-richer" phenomenon. 
However, the BA model could only capture the ``old-get-rich" phenomenon or the ``first-mover-advantage" whereby older nodes increase their connectivity and become dominant at the expense of younger nodes in the network.
It does not take into account the competitive characteristics of a node that help them flourish in a very short period of time \cite{Adamic00power-lawdistribution, Kong_2008}; for instance, in citation networks a few research papers are able to gain lot of citations within a short span of time \cite{AMANCIO2012427}. 
Using this as a motivation, Bianconi and  Barab{\'a}si~\cite{bianconi2001bose, Bianconi_2001} introduced a new class of network growth models in which the incoming nodes form connections based on inherent characteristics of nodes such as novelty, usefulness, etc., captured through a fitness value \cite{Nguyen2012, PhysRevLett.89.258702, PhysRevLett.93.188701}. This was inspired by the ``fit-get-richer" phenomenon observed in real-world networks~\cite{bianconi2001bose}. Following this, Ergun and Rodgers~\cite{ergun2002growing} analysed the degree distribution of network growth models with an attachment mechanism combining the degree and fitness information of nodes in an additive and multiplicative manner.


There are several real-world networks in which the aspect of {\em space} plays an important role to understand dynamics of network evolution. For instance, in the biological domain the regions in brain that are spatially closer have a higher probability of being connected as compared to the far-off regions \cite{Bullmore2009}. Similar significance of space can also be observed in online social networks capturing spatial features \cite{10.1145/2398776.2398793, 8368313}, transportation networks \cite{transport}, and communication networks. To model networks incorporating spatial features, the class of spatial network models has been proposed. A basic spatial model incorporating notions of preferential and spatial attachment was proposed by Yook et al. \cite{yook2002modeling} to capture underlying mechanisms driving the evolution of the Internet topology. Subsequently, Kaiser et al. \cite{Kaiser_2004} analysed a spatial growth model where the edge connection probability decreases with node distance either in an exponential or a power-law manner to explain multiple, interconnected clusters that emerge in real-world networks. See Barthélemy~\cite{Barth_lemy_2011} for a comprehensive study on spatial networks.
Recently, we  proposed a new spatial growth model \cite{mohapatra2020modeling} that was better able to capture the five growth patterns presented in \cite{chakraborty2015categorization}, compared to the preferential attachment model and its variants (i.e., additive and multiplicative fitness models \cite{PhysRevLett.89.258702,servedio2004vertex}).



In this paper, we extend our previous studies \cite{10.1007/978-3-319-55471-6_4,mohapatra2020modeling} and address the following problem statement: {\em Given an influential node with high visibility at a certain point in time, how would its visibility evolve over time}? We study this phenomenon across three popular and diverse network growth models -- Barab{\'a}si-Albert (BA) model, (additive, multiplicative, and general) fitness based model and spatial models.  
One of the primary theoretical findings that we continue to build on from our previous work \cite{10.1007/978-3-319-55471-6_4} is that leaders are able to gain more visibility in the multiplicative fitness model setting as compared to the BA and additive fitness models. Along with this, in the multiplicative fitness model, the influential nodes are able to increase their visibility over time, given that their fitness value remains high in comparison to the rest of the network. On the other hand, the visibility values always decrease over time for the BA and additive fitness models (see Section \ref{section:leader} for details). In Section~\ref{subsec:general-fitness}, we study a general framework of fitness models and observe that a non-linear attachment rule based on degree and fitness would lead to highly dominant nodes and make it exceedingly difficult for new nodes to gain influence in the network. Experimental analysis provided in Section \ref{sec:Simulation-Fitnessmodels} supports our theoretical analysis that suggest multiplicative fitness models best explain the prolonged influence of leader nodes in certain real-world networks, while at the same time allowing new high fitness nodes to gain influence over time. However, we observe that multiplicative fitness models do not allow multiple influential nodes to exist at the same time.
In Section~\ref{sec:Theory-SpatialModels}, we give theoretical insights on how spatial growth models can allow a multiplicity of leaders to coexist, while Section~\ref{sec:SimResults-SpatialModels} provides experimental justifications for the same.

\textbf{Reproducibility:} To encourage reproducible research, the codes are publicly available at \url{https://github.com/mittalshravika/Network-Growth-Models}.

\section{Network growth models}\label{section:models}
Following our previous work \cite{10.1007/978-3-319-55471-6_4}, here we set up the notations and problem definition. 
Consider the following sequence of graphs $\{\bG_t, \ t=0,1,\ldots \}$, where 
$\bG_t = (V_t,\bEg_t)$, with $V_t$ and $\bEg_t$ being the set of nodes and edges in $\bG_t$ respectively. In a network growth model, we have 
$V_t \subset V_{t+1}$ and $\bEg_t \subseteq \bEg_{t+1}$ 
for every $t=0,1,\ldots$. In other words, new nodes arrive at every time step $t$, and form 
connections with existing nodes, thus adding to the edge set of the previous graph $\bG_{t-1}$.
For purposes of simplicity, here
we consider the basic model where a single node enters at any time step $t$, 
and forms a connection with one node in the 
existing graph $\bG_{t-1}$. Therefore, we can label the incoming node by the time index of its entry to the network, which leads to $V_t = \{0,1,\ldots,t\}$ for $t = 0,1,\ldots$. Note that all our results can be easily extended to more general scenarios where multiple nodes can enter the network and incoming nodes can form multiple connections.
At time $t$, let the degree of the node $i$ in $V_t$ be denoted by $D_t(i)$. Also, let the rv $S_{t+1}$ denote the node with which an incoming node $t+1$ connects.

{\bf Barab{\'a}si-Albert (BA) model}:
In the preferential attachment mechanism \cite{barabasi1999emergence}, new nodes connect preferentially to existing 
nodes with higher degree. Let $\bp ^{BA}(t+1) = (p_i ^{BA}(t+1), \ i \in V_t)$ be the 
pmf with which the new node indexed as $t+1$ connects with the existing graph $\bG_t$, i.e., 
$p_i ^{BA}(t+1)$ is the probability with which node $t+1$ connects with an existing node $i$. This is given by:
\begin{equation}
p_i ^{BA}(t+1) = \bP{ S_{t+1} = i \ | \ \bG _t } = 
\frac{D_t(i)}{\sum _{j \in V_t} D_t(j)}, \ i \in V_t.
\label{eq:BA-visibility}
\end{equation}
Note that we term node $i$'s \emph{visibility} in the graph $\bG_t$ by $p_i ^{BA} (t+1)$.

{\bf Fitness based attachment rules}:
In fitness based models~\cite{bianconi2001bose,ergun2002growing,mitzen1}, 
every node is assumed to have a fitness value independently drawn 
from a distribution, and new nodes connect preferentially on the basis of the fitness and degree values of the existing nodes. We describe multiple ways in which such an attachment could occur. 

Assume a sequence of i.i.d. fitness rvs $(\xi, \xi_i, \ i=0,1,\ldots)$ with $\xi _i$ denoting the fitness value of node $i$.
A generic fitness model can be described by the following 
attachment rule
\begin{equation}
p_i ^{GF}(t+1) = \frac{g(\xi _i , D_t(i))}{\sum _{j \in V_t} g( \xi _j , D_t(j)) }, \ i \in V_t.
\label{eq:GF-visibility}
\end{equation}
for an attachment function $g: \R \times \R \to \R$ which determines the relative importance of 
the fitness and degree values. 
In the additive fitness model, new nodes connect preferentially to existing nodes having a higher sum of degree and fitness value. For $t=0,1,\ldots$, 
let the pmf delineating formation of new connections at time $t+1$ be given by $\bp^{AF}(t+1)$, where
\begin{equation}
p_i ^{AF}(t+1) = \frac{\xi _i +  D_t(i)}{\sum _{j \in V_t} (\xi _j +  D_t(j))}, \ i \in V_t.
\label{eq:AF-visibility}
\end{equation}
Similarly, the attachment rule for multiplicative fitness (MF) model is given by
\begin{equation}
p_i ^{MF}(t+1) = \frac{\xi _i \cdot  D_t(i)}{\sum _{j \in V_t} \xi _j \cdot  D_t(j)}, \ i \in V_t.
\label{eq:MF-visibility}
\end{equation} 
Therefore, the \emph{visibilities} of node $i$ in graph $\bG_t$ 
are given by $p_i ^{AF}(t+1)$ and $p_i ^{MF}(t+1)$ for the additive and multiplicative fitness models,
respectively. Note that the influential nodes as described in 
Section \ref{sec:Introduction} relate to nodes having high visibility as defined for the particular network growth model in question.

{\bf Spatial attachment rules:} 
In spatial models~\cite{yook2002modeling,Kaiser_2004,Ferretti_2011}, every node is assumed to have a location vector drawn from a distribution over a location space $A$. Assume a sequence of i.i.d. location rvs $(\chi, \ \chi _i , \ i=0,1,\ldots )$ and i.i.d. fitness rvs $(\xi, \xi _i, \ i=0,1,\ldots)$ with $\chi _i$ and $\xi _i$ denoting the location and fitness values of node $i$ respectively. A generic spatial attachment model can be given by the following attachment rule
\begin{equation}
    p_i ^{AT}(t+1) = \frac{ h(\chi _i, \chi _{t+1}; \xi _i, D_i(t)) }
    { \sum _{j \in V_t} h(\chi _j, \chi _{t+1}; \xi _j , D_j(t)) } .
\end{equation}

The attachment probability now depends on the location vector of the new 
node $\chi _{t}$. This is different from the models described previously. 
Therefore, we could have multiple definitions of visibility. A global 
variant of visibility is given below 
\[
p_i ^{\mbox{global}}(t+1) = \mathbb{E} _\chi 
\left[ 
\frac{h( \chi _i , \chi; \xi _i, D_i(t) )}
{ \sum _{j \in V_t} h( \chi _j , \chi; \xi _j , D_j(t) ) }
\right],
\]
while a local version is given as
\[
p_i ^{\mbox{local}} (t+1) = \frac{ h( \chi _i , \chi _i ; \xi _i, D_i(t) ) }
{ \sum _{j \in V_t} h( \chi _j , \chi _i ; \xi _j, D_j(t) ) },  
\]
where $h: A \times A \times \R \times \N \to \R$. We find it useful to consider the following 
separable form of attachment function 
\begin{equation}
h(\chi _1, \chi _2 ; \xi _1, D_1 ) = \alpha (\chi _1, \chi _2) \beta ( \xi _1, D_1), \ 
\chi _1, \chi _2 \in A, 
\label{eq:spatial-attachment-separable-form}
\end{equation}
where $\alpha: A \times A \to \R$ and $\beta : \R \times \N \to \R$. 
While the notion of global visibility models the overall attractivity of a node in the entire attribute space, the local visibility considers only its
attractivity from the local neighborhood in the attribute space of a node.
This model allows for nodes whose global attractivity is low, with their local attractivity being high.

\section{Analytical results on node visibility -- BA and Fitness models}\label{section:leader}

In this section, we study and compare the evolution of visibility of a node over time for the BA model and two fitness models, namely the additive and multiplicative fitness models. The following lemma describes the change in visibility with time for the three growth models.

First, we introduce some notation: Define $\Xi _t = \sum _{i \in V_t} \xi _i$
and $\psi _t = \sum _{i \in V_t} \xi _i D_t(i)$, for 
$t=0,1,\ldots$. 

\begin{lemma}
For every $t=0,1,\ldots,$ and $i$ in $V_{t-1}$: Let $\bG _{t-1}$ be the graph at time $t-1$,
 we have
\begin{itemize}
\item[(i)]
\begin{equation}
\bE{ p_i ^{BA} (t+1) - p_i ^{BA} (t) \ | \ \bG _{t-1} } = - \frac{D_{t-1}(i)}{4t(t-1)},
\label{eq:Change-BA-expected}
\end{equation}
\item[(ii)]
\begin{align}
&\bE{ p_i ^{AF} (t+1) - p_i ^{AF} (t) \ | \ \bG _{t-1}, \xi_t } 
\nonumber \\
& \hspace{2mm} = 
- \frac{\left( \xi _i +D_{t-1}(i) \right) \left( \xi _t + 1 \right)}
{ \left( \Xi _{t-1} + 2(t-1) \right) \left( \Xi_t + 2t \right) },
\label{eq:Change-AF-expected}
\end{align}
\item[(iii)]
\begin{align}
&\bE{ p_i ^{MF} (t+1) - p_i ^{MF} (t) \ | \ \bG _{t-1}, \xi _t } 
\nonumber \\
& \hspace{2mm} \gtrsim 
 \xi _i  D_{t-1} (i) \frac{ \sum _{j \neq i} \xi _j D_t(j) 
\left[ \xi _i - \xi_t - \xi_j \right] }
{ \psi _{t-1} ^2 (\psi _{t-1} +\xi _i + \xi_t)  } .
\label{eq:Change-MF-expected}
\end{align}
\end{itemize}
\label{lemma:Change_in_visibility}
\end{lemma}

\myproof
Fix $t=0,1,\ldots$, and $i$ in $V_t$.

\textbf{Preferential Attachment model:}
The difference in the visibility of node $i$ in the BA model between time $t+1$ and $t$ is given as
\begin{align}
    &p_i ^{BA}(t+1) - p_i^{BA}(t) = 
    \frac{D_t(i)}{2t} - \frac{D_{t-1}(i)}{2(t-1)}
    \nonumber \\
    & \hspace{2mm} = \frac{D_{t-1}(i)+\1{S_t = i}}{2t} - \frac{D_{t-1}(i)}{2(t-1)}
    \label{eq:proof-BA-Change-1}
\end{align}
The above follows by noting that the sum of degree rvs $D_t(i)$ for all the nodes in the vertex set $V_t$ equals $2t$. Furthermore, by noting that when looking at the expected difference in visibility conditioned on the graph at time $t-1$, $S_t$ is the only random variable in 
\eqref{eq:proof-BA-Change-1}, we obtain
\begin{align}
    &\bE{ p_i ^{BA} (t+1) - p_i ^{BA} (t) \ | \ \bG _{t-1} } 
    \nonumber \\
    &= \frac{D_{t-1}(i)+\bP{S_t = i \ | \ \bG_{t-1}}}{2t} - \frac{D_{t-1}(i)}{2(t-1)}
\end{align}
and \eqref{eq:Change-BA-expected} follows.

\textbf{Additive Fitness model:}
Similarly in the additive fitness model, the difference in the visibility of node $i$ can be 
written as
\begin{align}
    &p_i^{AF}(t+1) - p_i ^{AF}(t) 
    \nonumber \\
    &= \frac{\xi _i + D_t(i)}{ \sum _{j \in V_t} \xi _j + D_t(j)  }  - 
    \frac{\xi _i + D_{t-1}(i)}{ \sum _{j \in V_{t-1}} \xi _j + D_{t-1}(j)  }
    \nonumber \\
    &= \frac{\xi _i + D_{t-1}(i)+ \1{S_t = i} }{ \Xi_{t-1} + \xi _t + 2t  }  - 
    \frac{\xi _i + D_{t-1}(i)}{ \Xi_{t-1} + 2(t-1)  }.
    \label{eq:proof-AF-change}
\end{align}
Taking expectation on both sides
conditioned on $\bG_{t-1}$ and $\xi_t$ leads to \eqref{eq:Change-AF-expected}.

\textbf{Multiplicative Fitness model:}
The difference in the visibility of node $i$ can be written for the multiplicative model as follows
\begin{align}
    &p_i ^{MF}(t+1) - p_i ^{MF} (t) 
    \nonumber \\
    &= \frac{\xi _i D_t(i)}{\sum _{j \in V_t} \xi _j D_t(j)} - 
    \frac{\xi _i D_{t-1}(i)}{\sum _{j \in V_{t-1}} \xi _j D_{t-1}(j)}
    \nonumber \\
    &= \frac{\xi _i \left[ D_{t-1}(i) + \1{ S_t = i} \right]}{\psi _{t-1} + \xi_{S_t} + \xi _t} 
    - \frac{\xi _i D_{t-1}(i)}{\psi _{t-1}}
\end{align}
Furthermore, we lower bound the expected change 
in visibility as follows
\begin{align}
    &\bE{ p_i ^{MF}(t+1) - p_i ^{MF}(t) \ | \ \bG_{t-1}, \xi _t} 
    \nonumber \\
    &= \xi _i \left[ \frac{D_{t-1}(i)+1}{\psi _{t-1} +\xi_i + \xi _t} - \frac{D_{t-1}(i)}{\psi _{t-1} }  \right] \bP{S_t = i \ | \ \bG_{t-1}, \xi _t } \nonumber \\
    & \hspace{2mm} + \xi _i \sum _{\ell \neq i} \left[ \frac{D_{t-1}(i)}{\psi _{t-1} +\xi_\ell + \xi _t} - \frac{D_{t-1}(i)}{\psi _{t-1} }  \right] \bP{S_t = \ell \ | \ \bG_{t-1}, \xi _t } \nonumber \\
    & \approx \xi _i \Bigg[ \frac{\psi _{t-1} \bP{S_t = i \ | \ \bG_{t-1}, \xi _t } }
    { \psi _{t-1} \left( \psi _{t-1} + \xi _i + \xi _t \right) } - \frac{D_{t-1}(i)}{\psi _{t-1}}
    \nonumber \\
    & \hspace{4mm} \times \Bigg[ \sum  _{\ell \neq i }
    \bP{ S_t = \ell \ | \ \bG_{t-1}, \xi _t}  \cdot \frac{\xi _\ell + \xi _t}{ \psi _{t-1} + \xi _\ell + \xi_t }
    \Bigg] \Bigg]
    \nonumber \\
    & \geq \xi _i \Bigg[ \frac{ \xi _i D_{t-1}(i) }
    { \psi _{t-1} \left( \psi _{t-1} + \xi _i + \xi _t \right) } 
    \nonumber \\
    & \hspace{4mm}- \frac{D_{t-1}(i)}{\psi _{t-1}}  \cdot 
    \frac{\sum _{ \ell \neq i} \xi _\ell D_{t-1}(\ell) (\xi _\ell + \xi _t ) }{\psi _{t-1} ^2}
    \Bigg]
    \nonumber \\
    & \simeq \xi _i D_{t-1}(i)
    \left[ \frac{\xi _i \psi _{t-1}  -
    \sum _{ \ell \neq i} \xi _\ell D_{t-1} (\ell) (\xi _\ell +\xi _t) }{ \psi _{t-1} ^2 (\psi _{t-1} +\xi _i + \xi_t) }
    \right]   
    \nonumber
\end{align}
and the result follows.
\myendpf

We observe from \eqref{eq:proof-BA-Change-1} that a node’s visibility increases if it forms a new edge connection in the BA model. However, it is also evident from Lemma~\ref{lemma:Change_in_visibility} that the visibility of the node decreases in expectation, in a manner that is directly proportional to its degree $D_{t-1}(i)$. This can be understood from the fact that higher degree nodes in the network have higher visibility values as a result of which, their decrease in visibility would be more as compared to nodes that have lower visibility values. In the additive fitness model, we can infer from \eqref{eq:proof-AF-change} that a node’s visibility increases when it forms a new edge, provided $\Xi_{t-1} + 2(t-1) > (\xi _t + 2) [\xi _i + D_{t-1}(i) ]$. This condition is expected to hold for large values of $t$, unless the fitness value $\xi_t$, or $\xi _i$, or both, are very large. Similar to the BA model, node visibility decreases in expectation, with the magnitude of decrease being directly proportional to the sum of degree and fitness values, and the fitness value of the new incoming node $\xi _t$.

In contrast with the above, we can see from \eqref{eq:Change-MF-expected} that in expectation, the nodes are able to increase their visibility over time, given that their fitness value remains large with respect to the network. In addition to this, the expected change in visibility directly depends on the product of fitness and the present degree of the node, boosting the visibility of a leader much more as compared to the BA and additive fitness models. Note that we derive results for change in visibility values over a single time step. The results can be easily generalized to any fixed number of time steps.

\subsection{Node visibility over time -- General fitness model}
\label{subsec:general-fitness}

We study the change in node visibility over time for a general fitness model described in \eqref{eq:GF-visibility}. For ease of notation, we define $\Gamma _t = \sum _{i \in V_t} g( \xi _i, D_t (i) ) $ and $\Delta _{t,i} ^g = g(\xi _i, D_{t-1}(i) +1) - g(\xi _i, D_{t-1}(i))$ for $i$ in $V_t$ and $t=1,2,\ldots$.

\begin{lemma}
For every $t=0,1,\ldots,$ and $i$ in $V_{t-1}$: Let $\bG_{t-1}$ be the graph at time $t-1$, we have
\begin{align}
&\bE{ p_i ^{GF} (t+1) - p_i ^{GF} (t) \ | \ \bG _{t-1}, \xi _t }
\nonumber \\
& \simeq \left( 
\frac{g(\xi_i,D_{t-1}(i))}{ \Gamma _{t-1} ^2 }
\right)
\Bigg[ 
\frac{g(\xi_i,D_{t-1}(i))}{\Gamma _{t-1}} \cdot 
\left( \Delta _{t,i}  ^g - g(\xi _t,1) \right)
\nonumber \\
& \hspace{2mm}+ 
\sum _{k \neq i} 
\left(
\frac{g(\xi_k,D_{t-1}(k))}{\Gamma _{t-1}} \right)
\left( 
\Delta _{t,i} ^g -\Delta _{t,k} ^g - g(\xi_t,1)
\right) 
\Bigg].
\label{eq:Generalfitness-expectedvisibility}
\end{align}
\label{lemma:Change-in-visibility-GF}
\end{lemma}

\myproof

The difference in the visibility of node $i$ can be written as follows
\begin{align}
&p_i ^{GF}(t+1) - p_i ^{GF}(t) 
\nonumber \\
&= \mathbb{E} \Bigg[ \frac{g(\xi _i, D_{t-1}(i) + \1{S_t = i})}{\sum _{j \in V_t} g(\xi _j, D_{t-1}(j) + \1{S_t = j})}
\nonumber \\
& \hspace{5mm} -
\frac{g(\xi _i, D_{t-1}(i))}{\sum _{j \in V_{t-1}} g(\xi _j, D_{t-1}(j))} \Bigg]
\end{align}
We further introduce the following notation for $k$ in $V_t$, $\Omega _ {t,k} = \1{S_t = k}$.
\begin{align}
&p_i ^{GF}(t+1) - p_i ^{GF}(t) =
\nonumber \\
&= \Omega _{t,i} \left[  \frac{ g(\xi _i , D_{t-1}(i)+1)}
{
\Gamma _{t-1} + \Delta _{t,i} ^g
+ g(\xi _{t}, 1) 
}
- 
\frac{g(\xi _i, D_{t-1}(i))}{\Gamma _{t-1}}
\right]
\nonumber \\
&
+ \sum _{k \neq i}
\Omega _{t,k}
\left[ 
\frac{g(\xi _i , D_{t-1}(i))}
{ 
\Gamma _{t-1}
+
\Delta _{t,k} ^g
  + g(\xi _{t}, 1) }
- 
\frac{g(\xi _i, D_{t-1}(i))}{\Gamma _{t-1}}
\right]
\nonumber 
\end{align}
Furthermore, we introduce the following shorthand, $\hat{g} _{t,i} = g(\xi _i, D_{t-1}(i))$ and continue from above.
\begin{align}
&p_i ^{GF}(t+1) - p_i ^{GF}(t) 
\nonumber \\
&= 
\Omega _{t,i}
\
\left[ 
\frac{
\sum_{ \substack{ 
j \neq i \\
j \in V_{t-1} }}  \hat{g} _{t,j}
\Delta _{t,i} ^g
-g (\xi_t,1) \hat{g} _{t,i}
}
{
\Gamma _{t-1}
\left( 
\Gamma _{t-1}
+
\Delta _{t,i} ^g
 + g(\xi _{t}, 1) \right)
}
\right] 
\nonumber \\
&+
\sum _{k \neq i}
\Omega _{t,k}
\left[
\frac{ - \hat{g} _{t,i}
\Delta _{t,k} ^g
-
g(\xi _t,1) \hat{g} _{t,i}
}
{ 
\Gamma _{t-1}
\left(
\Gamma _{t-1}
+
\Delta _{t,k} ^g
 + g(\xi _{t}, 1) 
\right) 
 }
\right] 
\label{eq:GeneralFitness-difference-expression}
\end{align}
Using expression \eqref{eq:GeneralFitness-difference-expression}, we approximate
the expected change in visibility for sufficiently large values of $t$, as follows
\begin{align}
&\bE{ p_i ^{GF} (t+1) - p_i ^{GF} (t) \ | \ \bG _{t-1}, \xi _t }
\nonumber \\
&\simeq 
\bP{ S_t = i \ | \ \bG _{t-1}, \xi _t }
\left[ 
\frac{ \left(
\sum_{ \substack{ 
j \neq i \\
j \in V_{t-1} }}  \hat{g} _{t,j}
\Delta _{t,i} ^g \right)
-g (\xi_t,1) \hat{g} _{t,i}
}
{
\Gamma ^2 _{t-1}
}
\right]
\nonumber \\
&- 
\sum _{ k \neq i }
\bP{ S_t = k \ | \ \bG_{t-1}, \xi_t }
\left[
\frac{  \hat{g} _{t,i}
\Delta _{t,k} ^g
}
{ 
\Gamma ^2 _{t-1}
}
 + \frac{g (\xi_t,1) \hat{g} _{t,i}}{\Gamma _{t-1} ^2} \right]
\end{align}
on substituting the expressions for 
$ \left\{ \bP{S_t=\ell \ | \ \bG_{t-1} , \xi_t}, \ \ell \in V_{t-1} \right\}$,
we obtain
\begin{align}
& \simeq \frac{ \hat{g} _{t,i}
\Delta ^g _{t,i} }{ \Gamma _{t-1} ^2 }
- \hat{g} _{t,i}
\sum _{k \neq i} \frac{ \hat{g} _{t,k}  }{\Gamma _{t-1} } 
\frac{ \Delta ^g _{t,k} }{ \Gamma _{t-1} ^2}
- \frac{g(\xi_t,1) \hat{g}_{t,i}}{\Gamma _{t-1} ^2}
\nonumber \\
&= \left( 
\frac{ \hat{g} _{t,i} }{ \Gamma _{t-1} ^2 }
\right)
\left[ 
\Delta _{t,i} ^g - \sum _{k \neq i} \frac{ \hat{g} _{t,k}  }{\Gamma _{t-1}} 
\Delta _{t,k}  ^g
- g (\xi_t,1)
\right]
\nonumber \\
&=
\left( 
\frac{\hat{g} _{t,i}}{ \Gamma _{t-1} ^2 }
\right)
\Bigg[ 
\left(
\frac{ \hat{g} _{t,i} }{\Gamma _{t-1}} \right)
\left( \Delta _{t,i}  ^g - g(\xi _t,1) \right)
\nonumber \\
& \hspace{2mm}+ \sum _{k \neq i} 
\left(
\frac{ \hat{g} _{t,k} }{\Gamma _{t-1}} \right)
\left( 
\Delta _{t,i} ^g -\Delta _{t,k} ^g - g(\xi_t,1)
\right) 
\Bigg].
\nonumber 
\end{align}  
\myendpf
The expected change in visibility will be positive if 
\[
\Delta _{t,i} ^g \geq \Delta _{t,k} ^g, \ k \neq i, k \in V_{t-1}
\]
and $\Delta _{t,i} ^g \geq g(\xi _t,1)$.
While this is a sufficient condition, the expected change in visibility will be positive 
if node $i$ has significantly large visibility in the network. 
Observe that for the BA and additive fitness models, $\Delta _{t,i} ^g = 1$.
While for the BA model, the approximation is too crude, we obtain a decrease in
expected visibility for the additive model from \eqref{eq:Generalfitness-expectedvisibility}. 
For the MF model,
\[
\Delta _{t,i}  ^g = \xi _i ( D _{t-1} (i) +1) - \xi _i D_{t-1} (i) = \xi _i.
\]
Therefore, $\Delta _{t,i} ^g$ will be greater for nodes with higher fitness values. 
However, an incoming node with a high fitness value can lead to a decrease in the expected 
visibility of an influential node. This agrees with our findings in Lemma~\ref{lemma:Change_in_visibility}.
For attachment functions which combine the fitness and degree information in a nonlinear fashion, influential nodes can retain greater visibility for longer periods of time while being protected from decrease in visibility due to new nodes with high fitness values. 
For example, for nonlinear attachment rules like
 $g(\xi _i, D_t(i)) = \left( \xi _i D_t(i) \right)^2$, 
$\Delta ^g _{t,i} = \xi _i ^2 \left( 2 D_t(i) +1 \right)$, i.e., it is a function 
of both the fitness and degree values.
In this scenario, new nodes with high fitness values cannot lead to a decrease in expected visibility for influential nodes because of their low initial degree. 
However, nodes with low degree and high fitness values could experience a decrease in their visibility because only nodes with significantly large fitness and degree values can 
increase their visibility in an expected sense, which is not the case for the multiplicative model. This would lead to a great difficulty for new nodes with high fitness values to attain visibility in the network. Furthermore, it will become progressively  more difficult for new nodes to become visible in the network.

\section{Experimental results on node visibility -- Fitness models}
\label{sec:Simulation-Fitnessmodels}
In order to illustrate Lemmas~\ref{lemma:Change_in_visibility} and \ref{lemma:Change-in-visibility-GF} and depict the change in visibility of influential nodes, we carry out two sets of simulation experiments. We compare how the visibility of leaders or influential nodes change over time in the Barabasi-Albert (BA), additive fitness (AF), multiplicative fitness (MF) and general fitness (GF) models. Throughout, the fitness variable $\xi$ is taken to be Pareto distributed with parameter $\alpha_p$. 

For each given growth model and parameter value of the Pareto distribution, we generate a graph $\bG_{T_0} ^X$, where $X \in \{BA, AF, MF, GF \}$ and $T_0 = 10000$. We define $p_{(k)}(T_0;\bG _t ^X)$ to be the visibility of the node in graph $\bG _{t} ^X$ which had the $k$-highest visibility in graph $\bG _{T_0} ^X$. For each growth model, starting from $\bG _{T_0} ^X$, we generate $R$ realizations $\bG _{T} ^{X,(1)}, \bG _{T} ^{X,(2)}, \ldots \bG _{T} ^{X,(R)}$ with $T=100000$, which are mutually independent conditioned on $\bG _{T_0} ^X$. Since we are interested in the evolution of the visibility of influential nodes, for the purpose of this experiment we track the visibility of the top 50 nodes starting from $t=10000$ to $t=100000$. However, conditioned on the graph at time $T_0$, the visibility values are random variables; therefore, we average the visibility values across all the realizations at any given time. We define $\bar{p} _{(k)}(T_0;t;X) = \frac{1}{R} \sum _{r=1} ^R p_{(k)}(T_0;\bG _t ^{X,(r)})$ as the averaged visibility of the node at time $t$ ($t>T_0$) which had the k-highest visibility at time $T_0$ for growth model $X$.

In other words, we track $p _{(k)} \left( T_0; \bG _t ^{X, (r)} \right)$ for $k=1,2,\ldots,50$, $r=1,2,\ldots,R$, $t=10000,10001,\ldots,100000$, and $X \in \{BA, AF, MF, GF\}$. For large enough independent runs $R$, we expect $\bar{p} _{(k)}(T_0;t;X)$ to be a reasonable approximation of $\bE{p _{(k)} \left( T_0; \bG _t ^{X, (1)} \right) \big|  \bG _{T_0} ^X}$, which is the expected value of visibility at time $t$ for the node which had the $k$-highest visibility value at time $T_0$ conditioned on the graph at time $T_0$.

\begin{figure}[!th]
  \centering
  \includegraphics[width=0.9\columnwidth]{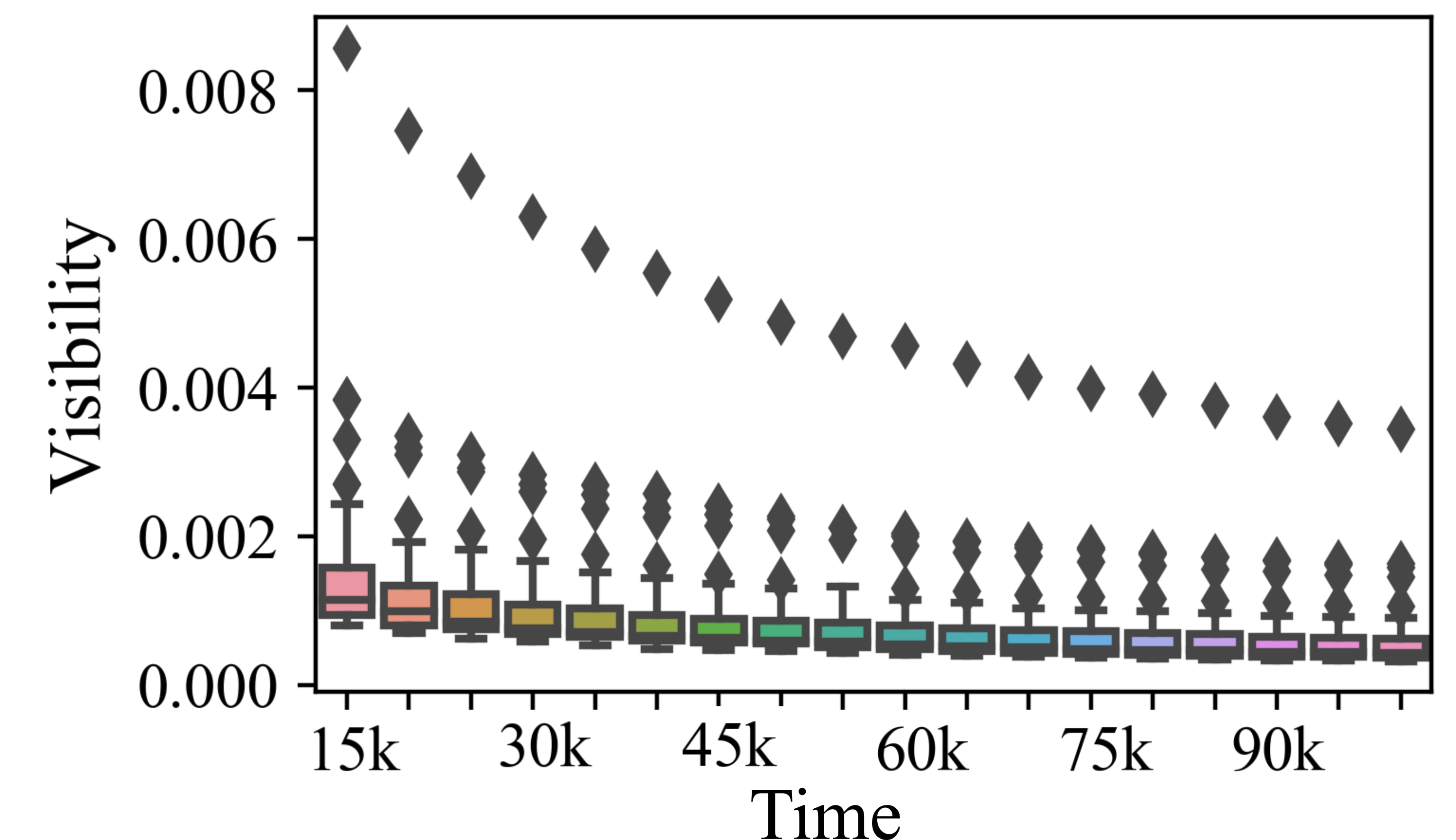}
  \caption{Visibility of nodes (averaged  over 50 independent runs) over time (after $T = 100000$ iterations) in the BA model.}
  \vspace{-5mm}
  \label{fig:BA_box_plot}
\end{figure}

\begin{figure*}[!th]
  \centering
  \includegraphics[width=0.9\textwidth]{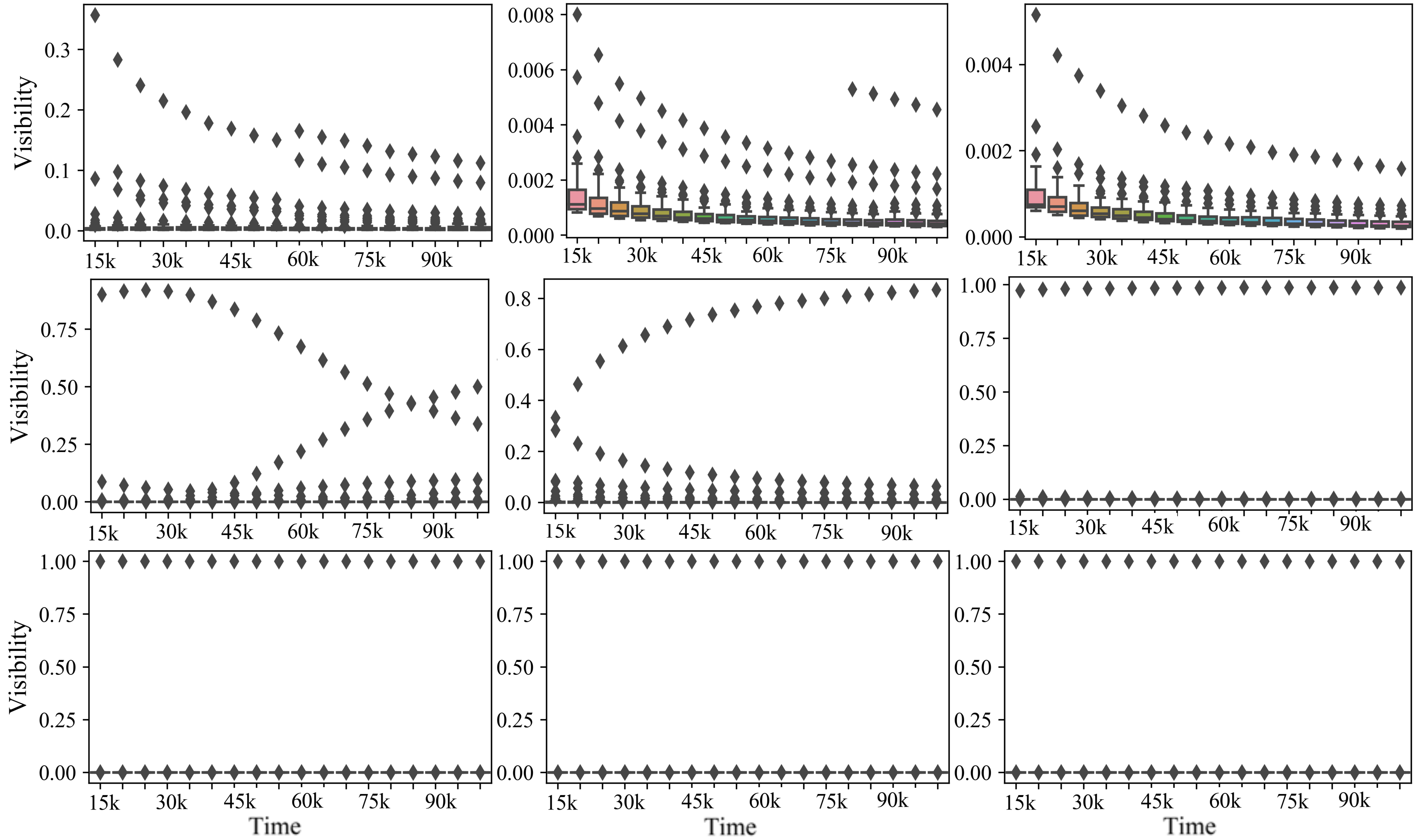}
  \caption{Visibility of nodes (averaged  over 50 independent runs) over time in three growth models. The columns represent different parameters of the Pareto distribution: $\alpha _p = 1, 2, 3$. Each row shows visibility results of the 50 most influential nodes after $T = 100000$ iterations for the three growth models -- additive, multiplicative and general fitness model.}
  \vspace{-5mm}
  \label{fig:box_plot}
\end{figure*}

Figures \ref{fig:BA_box_plot} and \ref{fig:box_plot} show the change in visibility of top 50 nodes at $t=T_0=10000$ when the graph is allowed to grow for 90000 iterations until $t=100000$. Visibility values averaged over $R=50$ independent runs from $T_0=10000$ are shown. We observe from the box plots in the two figures that the highest visibility nodes in the BA and AF slowly reduce in their visibility values over time as was predicted by Lemma~\ref{lemma:Change_in_visibility}. We also observe that lot more nodes in BA and AF models have significantly higher visibility values. However, for the multiplicative fitness model only 2 nodes exhibit high visibility values (for $\alpha_p=1,2$), with one node dominating the entire network at any point of time for most of the duration. In the MF model, for $\alpha_p =1,2$, we observe that a node with lower visibility replaces one with higher visibility between $t=T_0=10000$ and $t=100000$, because the lower visibility node joined the network later but with a significantly higher fitness value. For $\alpha_p =3$, we do not see this behavior because it is less likely that a node with a significantly high fitness value will enter the network. 

\begin{figure*}[!th]
  \centering
  \includegraphics[width=\textwidth]{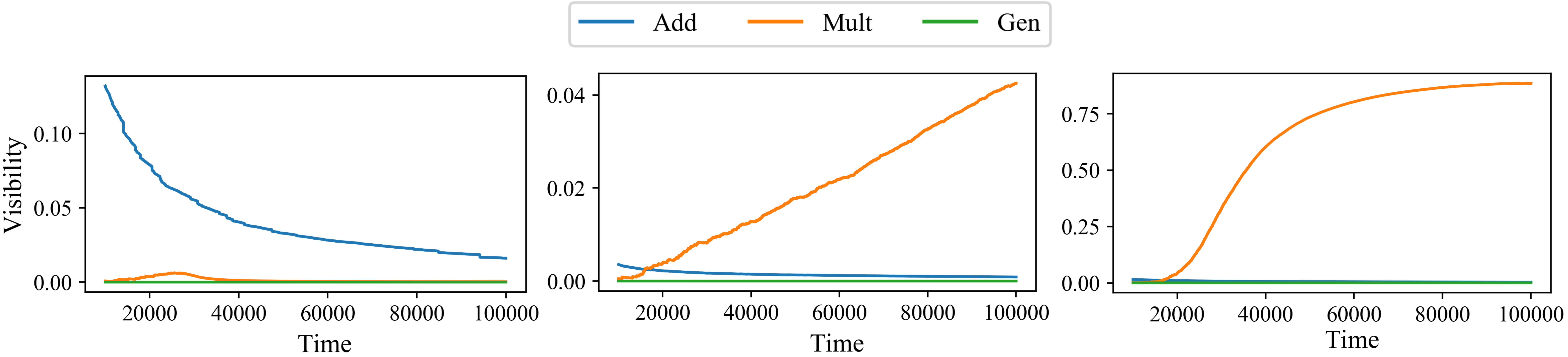}
  \caption{Visibility of new node (averaged  over 50 independent runs) added with a high fitness value over time in the three growth models -- additive, multiplicative and general fitness model. This represents results for different parameters of the Pareto distribution: $\alpha _p = 1, 2, 3$.}
  \vspace{-5mm}
  \label{fig:newnode}
\end{figure*}

While we observe that in both the MF and GF models, nodes with high visibility are able to maintain their influence (or, visibility) over time, we next investigate how easy it is for new nodes with high fitness to gain influence over time. For this purpose, we conduct an experiment where we introduce a node at time $t=T_0+1=10001$, with fitness value $\xi _{T_0+1} = 2 \max _{t \in V_{T_0}} \xi_t$, i.e., twice the fitness value of the maximum fitness value among all nodes until time $T_0$. For growth model $X \in \{AF, MF, GF \}$, we generate $R=50$ realizations beyond time $T_0$ by setting the fitness value of node $T_0+1$ as mentioned above. We define $p_i \left( \bG _t ^X \right)$ to be the visibility of the node $i$ in graph $\bG _t ^X$. We average the visibility values across $R$ realizations, $\bG _T ^{X, (1)}, \bG _T ^{X, (2)}, \ldots, \bG _T ^{X, (R)}$, and compute the averaged visibility of node $T_0+1$ defined as $\bar{p} _{T_0+1}(t) = \frac{1}{R} \sum _{r=1} ^R \left( \bG _t ^{X, (r)} \right)$, $t>T_0$. Subsequently, we track the visibility of this newly introduced node in the three fitness models and present the results in Figure~\ref{fig:newnode}. We observe that in the AF and GF models, the visibility of the newly introduced node decreases with time. This concurs with Lemma~\ref{lemma:Change_in_visibility} where we showed that the visibility of nodes in AF models decreases with time; and with Lemma~\ref{lemma:Change-in-visibility-GF} where we argue that in the general fitness model with a nonlinear attachment rule, it becomes progressively difficult for new nodes to get visible in the network. For the MF model with $\alpha_p = 1$, we observe that $\bar{p} _{T_0+1}(t)$ increases slightly but decays beyond $t=30000$. This is because nodes with even higher fitness values enter the network after node $T_0+1$. However, for $\alpha_p=2,3$ the visibility of node $T_0+1$ keeps increasing until $t=100000$. Note that for $\alpha_p=3$, node $T_0+1$ becomes dominant in the network very quickly because as the node increases its degree, the degree-fitness product becomes large compared to that of other nodes in the network because having a large fitness node is a rarer event for larger value of $\alpha_p$.

From the experiments we reaffirm that multiplicative models allow high visibility nodes to maintain their influence in the network for a longer period of time, while at the same time allowing high fitness nodes that are introduced later in the network to become influential. However, we observe that only a few number of nodes can be influential in the network at any given moment of time. This leads us to consider spatial attachment rules in conjunction with the multiplicative model to enforce regions of influence for each individual node such that multiple influential nodes can coexist in a network at the same time.

\section{Analytical results on node visibility -- Spatial models}
\label{sec:Theory-SpatialModels}
In the previous sections, we investigated the node visibility dynamics of the fitness models. A major takeaway was that multiplicative fitness models allow influential nodes to maintain their visibility while still permitting newly introduced nodes with high fitness to gain influence over time. However, a shortcoming of the MF model was that it could not support a multiplicity of influential nodes. We investigate whether a configuration of spatial models exists that retains the positive aspects of the MF model while addressing this shortcoming.

\subsection{Preliminaries - Results on various notions of visibility in spatial models}

Having defined global and local visibilities for a node in Section~\ref{section:models}, we find it helpful to define a notion of maximum visibility 
\begin{equation}
p_i ^{\mbox{max}} (t+1) = 
\max _\chi \frac{h(\chi _i , \chi; \xi _i, D_i(t))}
{ \sum _{j \in V_t} h(\chi _j, \chi ; \xi _j, D_j(t))}, 
\end{equation}
with $\chi _{t,i} ^*$ being the location vector for which the maximum is attained.
For rest of the analysis, we assume that the attachment function $h$ is separable into a product form of $ \alpha (\chi _1, \chi _2) \cdot \beta ( \xi, D) $ as shown in \eqref{eq:spatial-attachment-separable-form}. For analytical purposes, we set $\alpha(\chi _1, \chi_2) = e ^{-\gamma d(\chi _1, \chi _2)}$, where 
$d: A \times A \to \R$ is a metric on the location space $A$. 
\begin{lemma}    
For $\alpha (\chi _1,\chi _2) = e ^{-\gamma d(\chi _1,\chi_2)}$
and every $t=1,2,\ldots,$
\begin{equation}
\lim _{ \gamma \to \infty }
p_i ^{\mbox{local}} (t+1)  = 1 = 
\lim _{ \gamma \to \infty }
p_i ^{\mbox{max}} (t+1)
\label{eq:AttributeModel-Local-Max-Relation-1}
\end{equation}
and
\begin{equation}
\lim _{\gamma \to \infty} \chi _{t,i} ^* = \chi _i.
\label{eq:AttributeModel-Local-Max-Relation-2}
\end{equation}
In other words, for all $\delta >0$ and for all $t=1,2,\ldots$, there exists $\gamma _{t,\delta}$ such that 
for all $\gamma \geq \gamma _{t,\delta}$
\[
\left| p_i ^{\mbox{local}}(t+1) - p _i ^{\mbox{max}} (t+1) \right| < \delta
\]
and 
\[
\left| \chi _i - \chi _{t,i} ^* \right| < \delta. 
\]
\label{lemma:local_max_relationship}
\end{lemma}

\myproof
For $i$ in $V_t$ and $t=1,2,3,\ldots$
\begin{align}
&p_i ^{\mbox{local}} (t+1) 
\nonumber \\
&= 
\frac{\beta (\xi_i, D_i(t))}{ \beta ( \xi _i, D_i(t)) + \sum _{j \neq i} e ^{ - \gamma d (\chi _i, \chi _j) } \beta ( \xi _j, D_j(t)) }
\nonumber \\
& \geq \frac{ \beta (\xi_i, D_i(t)) }{  \beta( \xi_i,  D_i(t)) + \left( t \max _{k \in V_t} \xi _k \right)   \sum _{ j \neq i } e ^{ -t ^\epsilon d(\chi _i, \chi _j ) }   }
\nonumber \\
& \xrightarrow{\gamma = t ^\epsilon \to \infty} 1
\nonumber
\end{align}
and we obtain first part of \eqref{eq:AttributeModel-Local-Max-Relation-1}. Note that the scaling $\gamma = t ^\epsilon$ implies that for larger graphs, a smaller $\gamma$ would be necessary. Second part is obtained by noting that 
$ p_i ^{\mbox{max}} (t+1) \geq p_i ^{\mbox{local}}(t+1)$ for every node $i$ and time $t$. Equation \eqref{eq:AttributeModel-Local-Max-Relation-2} also follows similarly. 
\myendpf

Lemma \ref{lemma:local_max_relationship} suggests that $p_i^{\mbox{max}}$ is a good approximation for $p_i ^{\mbox{local}}$ 
when $\gamma$ is sufficiently large.
Next, we derive relationships between the global and local notions of visibility. 

\begin{lemma}
For $\epsilon >0$, we have
\begin{align}
&e ^{-2 \gamma \epsilon} \bP{ d(\chi, \chi_i) < \epsilon} p_i ^{\mbox{local}} (t+1) 
\leq 
p_i ^{\mbox{global}}(t+1) 
\nonumber \\
&\leq p_i ^{\mbox{max}}(t+1) \approx p_i ^{\mbox{local}}(t+1)
\end{align}
\label{lemma:Visibility-global-local-relationships}
\end{lemma}

\myproof
For convenience, we use the shorthand notation, $ \beta _{i}(t) = \beta(\xi _i, D_i(t)) $ for $i$ in $V_t$ and $t=1,2,\ldots$.
The upper bound on $p_i ^{\mbox{global}}(t+1)$ follows from the definition of $p_i ^{\mbox{max}}(t+1)$.
To obtain the lower bound, for a fixed $\epsilon>0$ we condition on the event $\1{d(\chi_i,\chi) < \epsilon}$
\begin{align}
& p_i ^{\mbox{global}}(t+1) 
\nonumber \\
&\geq 
\bP{ d(\chi_i, \chi) < \epsilon }
\nonumber \\
& \hspace{2mm} \times
\mathbb{E} _\chi 
\left[ \frac{h(\chi _i, \chi; \xi _i, D_i(t))}{ \sum _{j \in V_t} h(\chi _j, \chi;
\xi_j, D_j(t)) } \1{ d(\chi_i, \chi) < \epsilon }
\right]
\nonumber \\
& \geq 
\bP{ d(\chi_i, \chi) < \epsilon } \cdot
\min _{\chi: d(\chi, \chi _i) < \epsilon} \frac{h(\chi _i, \chi; \xi_i, D_i(t))}{ \sum _{j \in V_t} h(\chi _j, \chi; \xi_j, D_j(t)) }
\nonumber \\
&=
\bP{ d(\chi_i, \chi) < \epsilon } 
\nonumber \\
& \times 
\min _{\chi: d(\chi, \chi _i) < \epsilon} 
\frac{e^{-\gamma d(\chi, \chi _i)} \beta _i(t) }{ e^{-\gamma d(\chi, \chi _i)} \beta _i(t) + 
\sum _{j \neq i}    e^{-\gamma d(\chi, \chi _j)} \beta _j(t) }
\nonumber \\
& \geq 
\bP{ d(\chi_i, \chi) < \epsilon } \cdot
\frac{ e ^{-2 \gamma \epsilon}  \beta _i(t)}{ \beta _i(t) + 
\sum _{j \neq i}    e^{-\gamma d(\chi _i, \chi _j)} \beta _j(t) }
\nonumber \\
&= e ^{-2 \gamma \epsilon} \bP{ d(\chi, \chi_i) < \epsilon} p_i ^{\mbox{local}} (t+1)
\end{align}
where the penultimate step follows from triangle inequality.~\hfill\myendpf

Lemma \ref{lemma:Visibility-global-local-relationships} gives a lower and upper bound for the global visibility in terms 
of the local visibility. As argued previously, since the model is more concerned with local attractivity of nodes, we will present 
analysis for $p_i ^{\mbox{local}}$. Lemma \ref{lemma:Visibility-global-local-relationships} suggests that changes in 
the local visibility should also be aptly reflected in the global visibility.

\subsection{Node visibility over time}
\label{subsection:local_model}
 For convenience, we define some shorthand notation: For $i$ in $V_t$, $\beta _i(t) = \beta (\xi _i, D_i(t))$. For $i,j$ in $V_t$, $\hat{h}_{t,i \to j} = h(\chi _i, \chi _j; \xi _i, D_{t-1}(i))$, $\hat{h}_{t,i \to j} ^+ = h(\chi _i, \chi _j; \xi _i, D_{t-1}(i)+1)$ represent the attachment function for node $i$ at the location of node $j$, and $\Delta ^h _{t,i \to j} = h(\chi _i, \chi _j; \xi _i, D_{t-1}(i)+1) - h(\chi _i, \chi _j; \xi _i, D_{t-1}(i))$. For $k$ in $V_t$, $\Omega _{t,k} = \1{S_t=k}$ and $\Gamma _{t,k} = \sum _{i \in V_{t}} \hat{h}_{t,i \to k}$.

\begin{lemma}
For every $t=0,1,\ldots,$ and $i$ in $V_{t-1}$: Let $\bG_{t-1}$ be the graph at time $t-1$, we have
\begin{align}
&\bE{ p_i ^{\mbox{local}} (t+1) - p_i ^{\mbox{local}} (t) \ | \ \bG _{t-1}, \xi_t , \chi _t }
\nonumber \\
&\lesssim
C_1
\sum _{k \in V_t}  \beta _k (t-1) 
e ^{ - \gamma d(\chi_i, \chi _k)} 
\big\{ 
\xi _i  - e ^{ - \gamma d(\chi_i, \chi _k)}  \xi _k - \xi _t  \big\}
\label{eq:LocationBasedModel-LB-Lemma}
\end{align}
and 
\begin{align}
&\bE{ p_i ^{\mbox{local}} (t+1) - p_i ^{\mbox{local}} (t) \ | \ \bG _{t-1}, \xi _t, \chi_t}
\nonumber \\    
&\gtrsim
C_2 \sum _{k \in V_t}  \beta _k (t-1)
e ^{ - \gamma d(\chi_i, \chi _k)} 
\nonumber \\
& \hspace{2cm} \times
\big\{  \xi _i - 
 e ^{2 \epsilon \gamma} e ^{ - \gamma d(\chi_i, \chi _k)} \xi _k
-  e ^{3 \epsilon \gamma} \xi _t \big\}
\label{eq:LocationBasedModel-UB-Lemma}
\end{align}
with 
\[
C_1 = \frac{1}{ \left(  \sum _{j \in V_t} h( \chi _j, \chi _i; \xi _j, D_{t-1}(j) )^3 \right) }
\beta _i (t-1)
\]
and 
\begin{align}
C_2 &= \bP{ d(\chi _t, \chi) < \epsilon } e^{-\epsilon \gamma} 
\nonumber \\
& \times 
\frac{1}{ \left(  \sum _{j \in V_t} h( \chi _j, \chi _i; \xi _j, D_{t-1}(j) )^3 \right) }
\beta _i (t-1).
\nonumber 
\end{align}
\label{lemma:LocationModel-ChangeInVisibility}
\end{lemma}

\myproof
The difference in the local visibility of node $i$ can be written as follows
\begin{align}
&p_i ^{\mbox{local}}(t+1) - p_i ^{\mbox{local}}(t)
\nonumber \\
&= \frac{h(\chi _i, \chi _i; \xi _i, D_{t-1} (i) + \1{S_t = i})}
{ \sum _{j \in V_t} h(\chi _j, \chi _i; \xi _j, D_{t-1} (j) + \1{S_t = j})}
\nonumber \\
& \hspace{2mm} -
\frac{h(\chi _i, \chi _i; \xi _i, D_{t-1} (i))}
{ \sum _{j \in V_{t-1}} h(\chi _j, \chi _i; \xi _j, D_{t-1} (j))}
\nonumber \\
&= \Omega _{t,i}
\left[ 
\frac{\hat{h} ^+ _{t,i \to i}}
{ \Delta ^h _{t,i \to i} + \Gamma _{t,i}  + h(\chi _t, \chi _i; \xi_t, 1)} 
-
\frac{\hat{h} _{t,i \to i}}
{ \Gamma _{t,i}}
\right]
\nonumber \\
& + \sum _{k \neq i}
\Omega _{t,k}
\left[ 
\frac{ \hat{h} _{t,i \to i} }
{ \Delta ^h _{t,k \to i} + \Gamma _{t,i}  + h(\chi _t, \chi _i; \xi _t, 1)}
-
\frac{ \hat{h} _{t,i \to i}}
{  \Gamma _{t,i}  }
\right]
\nonumber \\
&\approx \Omega _{t,i} \left[
\frac{  \Delta ^h _{t,i \to i}   }{ \Gamma _{t,i} }
\right]  
\nonumber \\
& \hspace{2mm} -
\sum _{k \neq i}
\Omega _{t,k} 
\Bigg[ 
\frac{  \hat{h} _{t,i \to i} \Delta ^h _{t,k \to i}}
{ \Gamma _{t,i} ^2 } +\frac{h(\chi _t, \chi _i;\xi _t, 1) \hat{h} _{t,i \to i} }
{ \Gamma _{t,i} ^2 }
\Bigg]
\label{eq:AttributeModel-ChangeInVisibility}
\end{align}
Using \eqref{eq:AttributeModel-ChangeInVisibility}, we upper bound the expected change in local visibility 
by noting the fact that the expected increase in visibility is the most when the new node has the highest probability to form connection with node $i$, which occurs when the 
attribute of the new node is close to $\chi _i$
\begin{align}
&\bE{ p_i ^{\mbox{local}}(t+1) -  p_i ^{\mbox{local}}(t)  \ |  \bG _t, \xi_t, \chi _t }
\nonumber \\
& \leq 
\bE{  p_i ^{\mbox{local}}(t+1) -  p_i ^{\mbox{local}}(t)  \ |  \bG _t, \xi _t, \chi _t = \chi _i }
\nonumber \\
&\leq \frac{1}{\Gamma _{t,i} }
\Bigg[
\frac{ \hat{h} _{t,i \to i} }{ \Gamma _{t,i} } \Delta ^h _{t,i \to i}
-
\sum _{k \neq i} 
\left( 
\frac{ \hat{h} _{t,k \to i} }{  \Gamma _{t,i} }
\right)
\Bigg( \frac{ \hat{h} _{t,i \to i} }{ \Gamma _{t,i} } \Delta ^h _{t,k \to i} 
\nonumber \\
&
\hspace{2mm}
-
\frac{h(\chi _t, \chi _i; \xi _t, 1)  \hat{h} _{t,i \to i} }{ \Gamma _{t,i} } \Bigg)
\Bigg] 
\nonumber \\
& \approx
\frac{1}{ \Gamma _{t,i} ^3  }
\sum _{k \in V_t} 
\Big\{
\hat{h} _{t,i \to i} \hat{h} _{t,k \to i} \Delta _{t,i \to i} ^h
-
\hat{h} _{t,i \to i} \hat{h} _{t,k \to i} \Delta _{t,k \to i} ^h
\nonumber \\
&
\hspace{2mm}
-
\hat{h} _{t, k \to i} \hat{h} _{t, i \to i}
h(\chi _i, \chi _i ; \xi _t,  1)
\Big\}
\nonumber \\
& \approx
\frac{1}{ \Gamma _{t,i} ^3 }
\beta (\xi _i, D_{t-1}(i))\sum _{k \in V_t}  \beta (\xi _k, D_{t-1}(k)) 
e ^{ - \gamma d(\chi_i, \chi _k)} 
\times 
\nonumber \\
& \hspace{2cm}
\big\{ 
\Delta ^h _{t,i \to i} - \Delta ^h _{t, k \to i} 
-   \beta(\xi _t, 1) \big\}
\label{eq:ChangeInVisibility-UpperBound-FinalExpression}
 \end{align}
Observe that 
$\Delta ^h _{t, i \to i}= \beta(\xi _i, D_{t-1}(i)+1) - \beta(\xi _i, D_{t-1}(i))$
which equals $\xi_i$ in a multiplicative $\alpha$ model. Similarly, $\Delta ^h _{t, k \to i} $ equals $e^{-\gamma d(\chi_i, \chi _k)} \xi_k$. Therefore, in a multiplicative model, \eqref{eq:ChangeInVisibility-UpperBound-FinalExpression} reduces to 
\begin{align}
&\bE{p_i ^{\mbox{local}}(t+1) - p_i ^{\mbox{local}}(t) \ | \ \bG _t, \xi _t, \chi _t}
\lesssim
\frac{1}{ \Gamma _{t,i} ^3 }
\beta _i (t-1)
\nonumber \\
& \hspace{2mm}
\times 
\sum _{k \in V_t}  \beta _k ( t-1) 
e ^{ - \gamma d(\chi_i, \chi _k)} 
\big\{ 
\xi _i  - e^{-\gamma d(\chi_i, \chi _k)} \xi _k - \xi _t  \big\}
\end{align}
which gives the upper bound \eqref{eq:LocationBasedModel-UB-Lemma}. To lower bound the same, we define the following notation --
For $\epsilon >0$, 
\[
\chi _{i,\min} ^\epsilon = \arg \min _{\chi: d(\chi, \chi _i) < \epsilon} 
\frac{h(\chi _i, \chi; \xi_i, D_i(t))}
{ \sum _{j \in V_t} h(\chi _j, \chi; \xi_j, D_j(t))}.
\]
In other words, $\chi _{i,\min} ^\epsilon$ is the attribute vector on the 
$\epsilon-$ball around $\chi _i$ where the attachment probability to $i$ is the lowest. 

Using   \eqref{eq:AttributeModel-ChangeInVisibility}, we lower bound the expected change in local visibility by noting that conditioned on the event that the new node has an attribute vector within an $\epsilon-$ball around $\chi _i$, it will be minimum when it is equal to $\chi _{i,\min} ^\epsilon$. Accordingly, we define the following notation: $\hat{h} ^{\min} _{t,k \to i} = h(\chi _k, \chi _{i,min} ^{\epsilon} ; \xi _k, D_{t-1}(k))$. We lower bound using conditioning arguments
\begin{align}
&\bE{ p_i ^{\mbox{local}}(t+1) -  p_i ^{\mbox{local}}(t)  \ |  \bG _t, \xi _t, \chi _t }
\nonumber \\
& \geq 
\bE{  p_i ^{\mbox{local}}(t+1) -  p_i ^{\mbox{local}}(t)  \ \big| \ \bG _t, \xi _t,  
d(\chi _t, \chi) < \epsilon }
\nonumber \\
& \hspace{2mm} \times
\bP{ d(\chi _t, \chi) < \epsilon }
\nonumber \\
& \geq 
\bE{  p_i ^{\mbox{local}}(t+1) -  p_i ^{\mbox{local}}(t)  \ \big| \ \bG _t, \xi _t, 
\chi _t = \chi _{\min} ^\epsilon } 
\nonumber \\
& \hspace{2mm} \times \bP{ d(\chi _t, \chi) < \epsilon }
\nonumber \\
& \gtrsim
\bP{ d(\chi _t, \chi) < \epsilon }
\cdot
\frac{1}{ \Gamma _{t,i} ^3 }
\sum _{k \in V_t} 
\Big\{
\hat{h} ^{\min} _{t,i \to i} \hat{h} _{t,k \to i}
 \Delta ^h _{t, i \to i} 
\nonumber \\
&
- 
\hat{h} _{t, i \to i} \hat{h} ^{\min} _{t, k \to i} \Delta ^h _{t, k \to i}
-
\hat{h} ^{\min} _{t, k \to i} \hat{h} _{t, i \to i}
h(\chi _i, \chi _{i,\min} ^\epsilon ; \xi _t,  1)
\Big\}
\nonumber \\
& \gtrsim
\bP{ d(\chi _t, \chi) < \epsilon }
\frac{1}{ \Gamma _{t,i} ^3 }
\beta (\xi _i, D_{t-1}(i))
\nonumber \\
&
\times 
\sum _{k \in V_t}  \beta _k (t-1)  
e ^{ - \gamma d(\chi_i, \chi _k)} 
\nonumber \\
& \times 
\big\{ e^{-\epsilon \gamma} \xi _i - 
 e ^{\epsilon \gamma} e ^{ - \gamma d(\chi_i, \chi _k)} \xi _k
-  e ^{2 \epsilon \gamma} \xi _t \big\}
\label{eq:ChangeInVisibility-LowerBound-FinalExpression}
\end{align}
\myendpf
where the final step follows by applying triangle inequality.
From the lower bound \eqref{eq:LocationBasedModel-LB-Lemma} 
it is clear that the visibility will decrease when 
nodes close to $i$ have high fitness values or the new node has a very high fitness value. 
The upper bound \eqref{eq:LocationBasedModel-UB-Lemma} indicates that if the fitness of node $i$ 
is sufficiently high compared to other nodes in its local neighborhood, then its visibility should increase. This shows how the local neighborhood of particular node impacts its visibility. Also, given the local nature of the behavior of visibility, more nodes end up being visible in the spatial model compared to the multiplicative fitness model.

\begin{figure*}[!ht]
  \centering
  \includegraphics[width=0.9\textwidth]{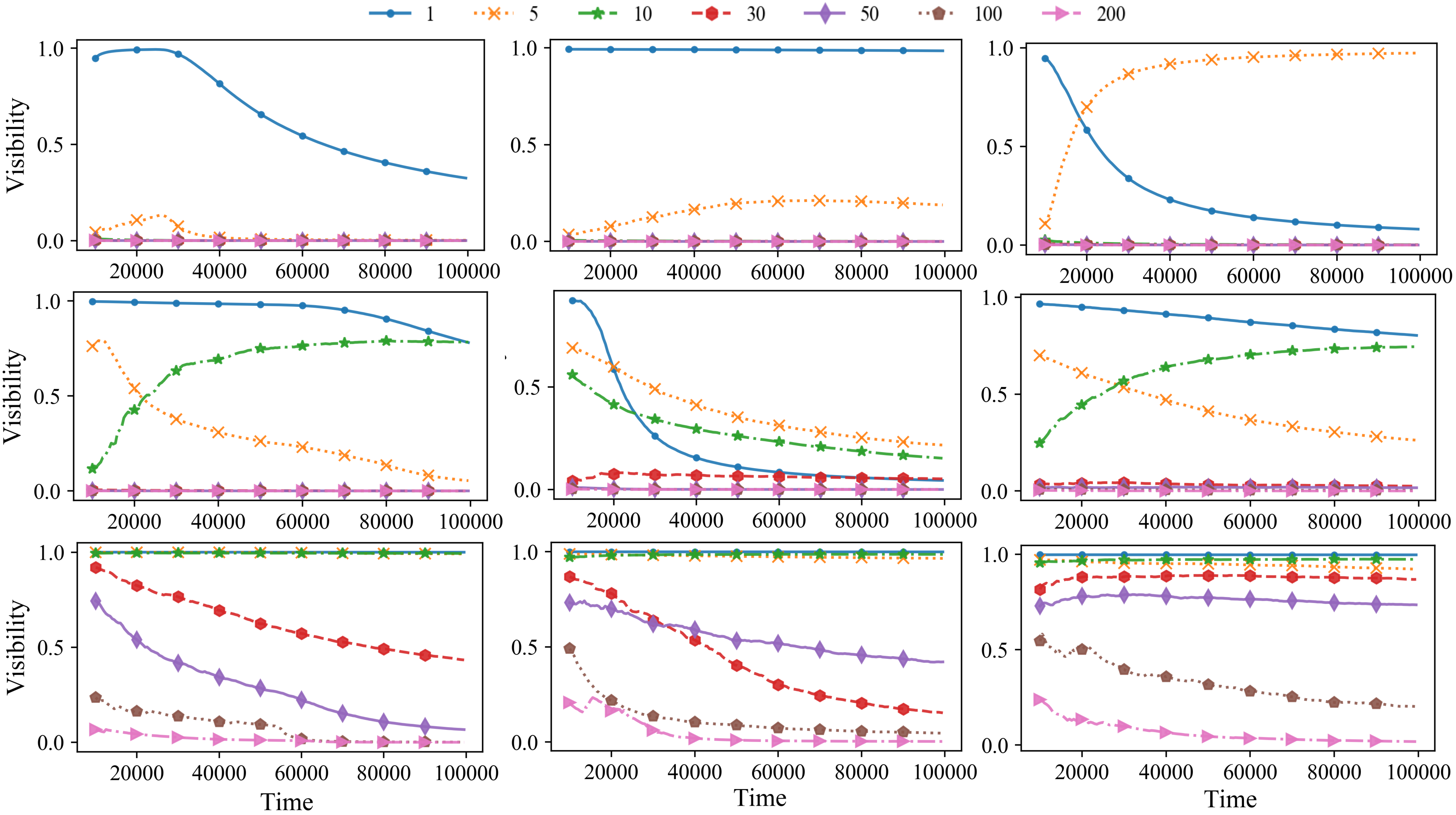}
  \caption{Visibility of nodes (averaged  over 50 independent runs) over time in the spatial growth model. The columns represent different parameters of the Pareto distribution: $\alpha = 1, 2, 3$. Each row shows visibility results for different values of $\gamma$ parameter, 5, 10 and 50 (from top to bottom).}
  \vspace{-5mm}
  \label{fig:local}
\end{figure*}

\section{Experimental results on node visibilities -- Spatial models}
\label{sec:SimResults-SpatialModels}
To illustrate the findings of Section \ref{subsection:local_model}, we perform experiments discussed in this section. In the previous section, we theoretically argued how the spatial (S) model with multiplicative $\beta$ would lead to multiple leaders coexisting in the network. Here we present experimental results that show the multiplicity of leaders that can coexist in the network and how this varies with the decay parameter $\gamma$. Throughout, the fitness variable $\xi$ is taken to be Pareto distributed with parameter $\alpha _p$.

For each $\alpha _p = 1,2,3$ and $\gamma = 5,10,50$, we generate a graph $\bG_{T_0} ^{S}$ as an instantiation of the spatial model, with $T_0 = 10000$. We denote $p ^{\mbox{local}}_{(k)}(T_0;\bG _t ^{S})$ to be the local visibility of the node in graph $\bG _{t} ^{S}$ which had the $k$-highest local visibility in graph $\bG _{T_0} ^{S}$.  As previously, we generate $R$ realizations $\bG _{T} ^{S,(1)}, \bG _{T} ^{S,(2)}, \ldots \bG _{T} ^{S,(R)}$ with $T=100000$, which are mutually independent conditioned on $\bG _{T_0} ^{S}$. We average the local visibility values across all the realizations at any given time and denote $\bar{p} ^{\mbox{local}} _{(k)}(T_0;t;S) = \frac{1}{R} \sum _{r=1} ^R p ^{\mbox{local}} _{(k)}(T_0;\bG _t ^{S,(r)})$ as the averaged local visibility of the node at time $t$ which had the k-highest visibility at time $T_0$ for the spatial model.

In other words, we track $p ^{\mbox{local}} _{(k)} \left( T_0; \bG _t ^{S, (r)} \right)$ for nodes $k=1, 5, 10, 30, 50, 100, 200$, runs $r=1,2,\ldots,R$, and decay parameters $\gamma = 5, 10, 50$, with $10000 \leq t \leq 100000$.

Figure \ref{fig:local} shows the change in local visibility of top $k^{\text{th}}$ nodes at $t=T_0=10000$ when the graph is allowed to grow for 90000 iterations until $t=100000$. Visibility values averaged over $R=50$ independent runs from $T_0=10000$ are shown. We observe that with increasing $\gamma$, the number of nodes with high local visibility increases in the network. This corroborates the insight that with increasing $\gamma$, the region of influence of nodes decreases leading to the potential of larger number of influential nodes in the network. For $\gamma = 5$ and $\alpha _p = 1$, we see that the $k=5^{\text{th}}$ node increases slightly in visibility beyond $t=10000$ but then decays beyond a certain point; and with $\alpha_p = 2$, the same node maintains its visibility until $t=100000$, while for $\alpha _p = 3$ the node increases its local visibility and dominates its region eventually. However for $k=10,30,...$, the corresponding nodes have very low values of local visibility. This changes for the $\gamma = 10$ case. Here, the $k= 10^{\text{th}}$ node also shows a high value of local visibility due to the reduced region of influence of nearby influential nodes. This becomes even more pronounced in $\gamma =50$ where we observe that even the $k=200^{\text{th}}$ node has non-trivial local visibility values that are maintained over some period of time for $\alpha _p =2, 3$. This shows that with increasing the decay parameter of the spatial model we can significantly increase the number of leaders in the network, many of whom can maintain their influence in their region.


\section{Conclusion}

In this paper, we studied the visibility profile of nodes in different classes of network growth models.
Firstly, we observed that in the multiplicative fitness model, nodes with high fitness values can successfully maintain visibility in the network to a greater extent when compared with the additive fitness and BA models. A general fitness model that has a non-linear attachment rule, e.g., that combines the degree and fitness values in a non-linear (quadratic) fashion, would also allow influential nodes to maintain visibility. However, unlike in multiplicative models, in these general fitness models with non-linear attachment rules we showed that it becomes progressively more difficult for new nodes with high fitness values to become influential in the entire network. We demonstrated through experimental results that in multiplicative models only a few number of nodes can be influential in the network at any given moment of time. This leads us to investigate spatial models that allows a multiplicity of influential nodes to exist in the network. We also show how the decay parameter in these spatial models can be used to control the number of leaders in the network. 

\section{Acknowledgements}
The authors would like to thank Dr. Ralucca Gera at Naval Postgraduate School, and Dr. Soham De at DeepMind, for insightful discussions and collaboration on previous works that led to this paper.


{\small
\bibliographystyle{plain}
\bibliography{references}}
\newpage
\vspace{-10mm}
\begin{IEEEbiography}[{\includegraphics[width=1in,height=1.25in,clip,keepaspectratio]{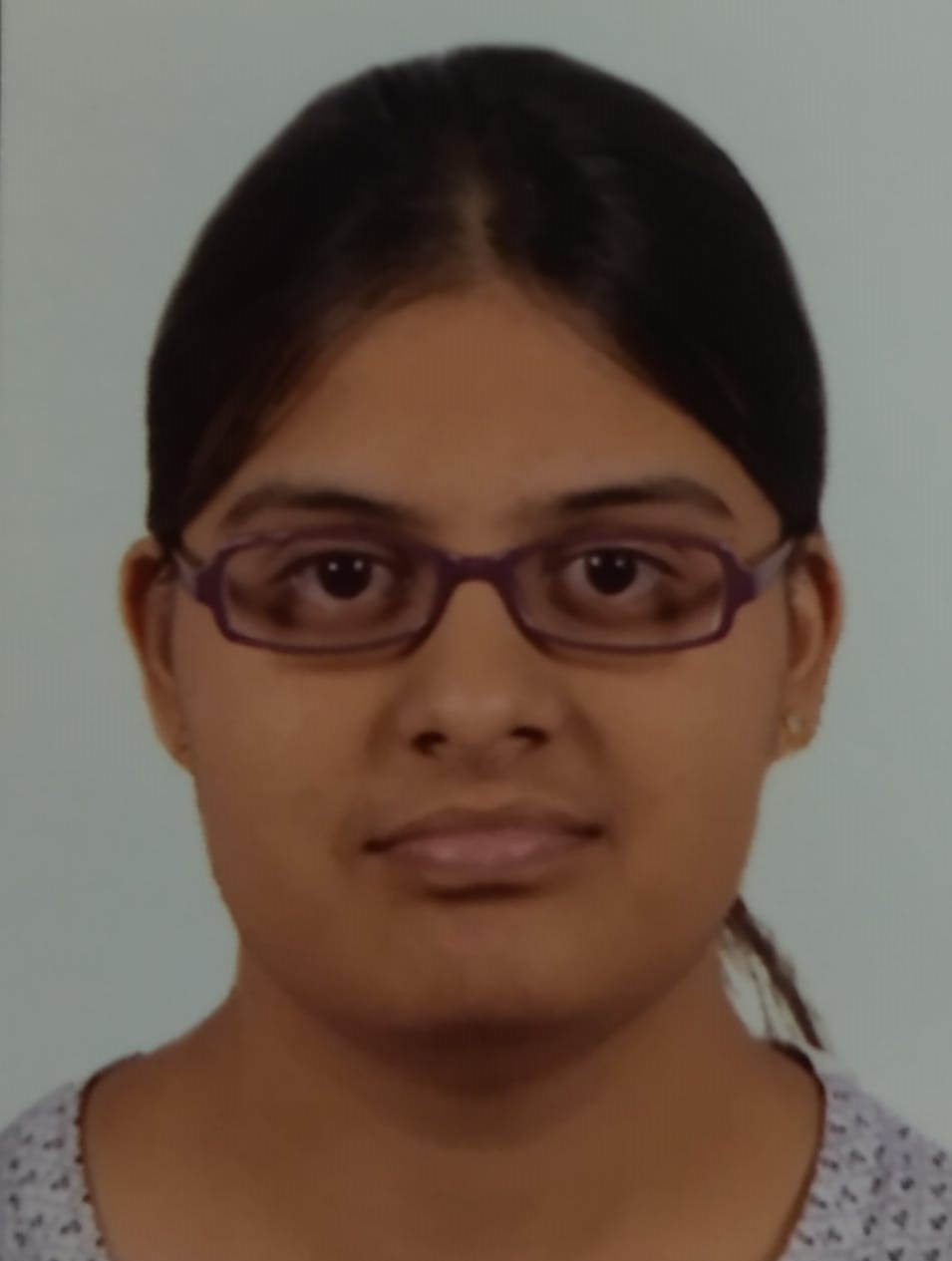}}]{Shravika Mittal} is a senior undergraduate student in Computer Science and Engineering at IIIT-Delhi. Her research interests include Network Science, Natural Language Processing and Machine Learning. She has received the Dean's list for Excellence in Academics, and Innovation in Research and Development. 
\end{IEEEbiography}

\vspace{-5mm}

\begin{IEEEbiography}[{\includegraphics[width=1in,height=1.25in,clip,keepaspectratio]{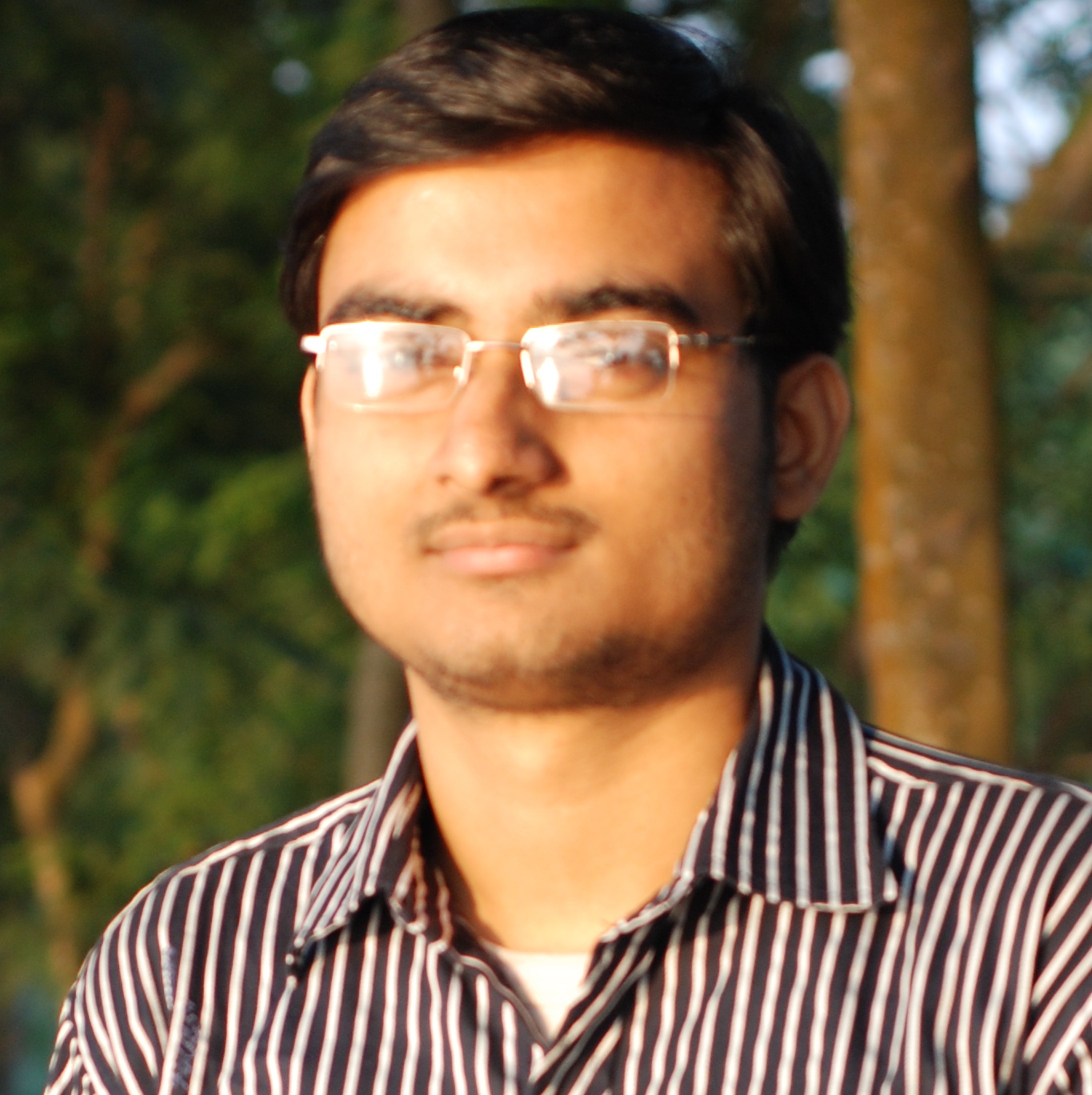}}]{Tanmoy Chakraborty} is an Assistant Professor and a Ramanujan Fellow at the Dept. of CSE, IIIT-Delhi, India, where he leads a research group, called LCS2 (\url{http://lcs2.iiitd.edu.in/}). His primary research interests include Social Network Analysis, Data Mining, and Natural Language Processing. He has received several awards including  Faculty Awards from Google, IBM and Accenture; Early Career Research Award, DAAD Faculty Fellowship. He is a member of ACM and IEEE. More details at \url{http://faculty.iiitd.ac.in/~tanmoy/}.
\end{IEEEbiography}

\vspace{-5mm}
\begin{IEEEbiography}[{\includegraphics[width=1in,height=1.25in,clip,keepaspectratio]{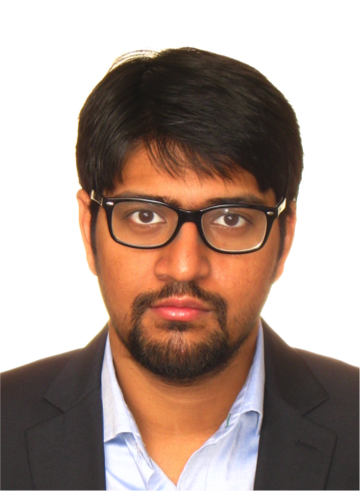}}]{Siddharth Pal}
received his Bachelors degree in Electronics and
Telecommunication Engineering from Jadavpur University, India, in
2011, and his Masters and Ph.D degrees both in Electrical Engineering
from University of Maryland College Park, USA, in 2014 and 2015 respectively. Since then he has been working as a research scientist
at Raytheon BBN Technologies. His research interests include network
science and analysis, game theory, machine learning with emphasis on
neural network based approaches, and stochastic systems.
\end{IEEEbiography}

\end{document}